\documentclass[times]{aastex62}

\usepackage{amsmath}  
\usepackage{amsfonts} 
\usepackage{amssymb}

\def\kms{\,{\rm km\ s}^{-1}}
\def\half{\tfrac{1}{2}}

\shorttitle{Gas in a box}
\shortauthors{Tremaine, Kocsis and Loeb}

\begin{document}

\title{Thermal equilibrium of an ideal gas in a free-floating box}

\author{Scott Tremaine}
\affiliation{Institute for Advanced Study, Princeton, NJ 08540, USA}
\affiliation{Canadian Institute for Theoretical Astrophysics, University of Toronto, 60 St.\ George Street,
  Toronto, ON M5S 3H8, Canada}

\author{Bence Kocsis}
\affiliation{Rudolf Peierls Centre for Theoretical Physics, Clarendon Laboratory, Parks Road, Oxford OX1 3PU, UK}

\author{Abraham Loeb}
\affiliation{Astronomy Department, Harvard University, 60 Garden St., Cambridge, MA 02138, USA}

\begin{abstract}

The equilibrium and fluctuations of an ideal gas in a rigid container
are studied by every student of statistical mechanics. Here we study
the less well-known case when the box is floating freely; in
particular we determine the fluctuations of the box in velocity and
position due to interactions with the gas it contains. This system is
a toy model for the fluctuations in velocity and position of a black
hole surrounded by stars at the center of a galaxy. These fluctuations
may be observable in nearby galaxies. 

\end{abstract}

\section{Introduction}

The determination of the equilibrium state of a dilute gas in a box is
perhaps the most famous problem in statistical mechanics, and usually
the first one to which students are exposed \citep{llsp}. The problem is
usually solved with the assumption that the box is fixed, or at least
very massive compared to the gas it contains. The purpose of this
short paper is to describe the equilibrium state when the box is free
to move and the mass of the box is comparable to the mass of the gas.

This system offers both a simple analytic problem in statistical
mechanics and a toy model for a more complicated, and still poorly
understood, problem from astrophysics: the Brownian motion of the
supermassive black holes found at the centers of galaxies due to the
gravitational forces from the surrounding stars (see \S4 for more
detail).

We consider an isolated rigid box of mass $M$ containing $N$ particles
of mass $m$, so the total gas mass is $M_g=Nm$. The particles collide
via short-range forces and we neglect the self-gravity of the gas. The
box is freely floating and there is a heater attached to the box that
maintains its temperature at a fixed value $T$ (thus we are working
with a canonical ensemble in which the ``heat bath" exchanges energy
but not momentum with the system; we could instead have used an
insulating box and worked with the microcanonical ensemble, with very
similar results). The center of mass of the system (box plus gas) is
at rest. We can assume that the box is cubical, with its symmetry axes
parallel to the coordinate axes $(x,y,z)$. Then the dynamics along
each axis is independent. What is the root-mean-square velocity of the
box along each axis?

There are two simple arguments leading to different answers, and both
turn out to be correct only in limiting cases.

First, we can argue that the box is in thermal equilibrium with the
gas. The mean-square velocity of the gas particles in one
dimension---say along the $x$-axis---is $\langle v_x^2\rangle=k_BT/m$
where $k_B$ is Boltzmann's constant. Then equipartition implies that
the mean-square velocity of the box along the $x$-axis should be
\begin{equation}
\langle V_x^2\rangle=\frac{k_BT}{M}.
\label{eq:en}
\end{equation}

The second argument starts with the observation that the center of
mass of the system is at rest so momentum conservation implies that
$MV_x=-m\sum_{i=1}^Nv_{xi}$ where $v_{xi}$ is the $x$-velocity of
particle $i$ and the sum is over all the particles. From the central
limit theorem, the mean-square value of the right-hand side should be
$m^2 N \langle v_x^2\rangle=mNk_BT=M_gk_BT$ so
\begin{equation}
\langle V_x^2\rangle = \frac{M_g}{M^2}k_BT,
\label{eq:mom}
\end{equation}
obviously not the same as (\ref{eq:en}).

\section{The mean-square velocity of the free-floating box}

Each microstate of the system we are examining is specified by the $N$
particle velocities $v_{xi}$ and the box velocity $V_x$, both measured
along the $x$-axis. Momentum conservation restricts the phase space to
the hyperplane $mv_{x1}+\cdots+ mv_{xN}+MV_x=0$.  The canonical
ensemble assigns to each microstate a probability proportional to
$\exp[-E/(k_BT)]$, where the kinetic energy
$E=\half m(v_{x1}^2+\cdots+ v_{xN}^2)+\half MV_x^2$. Therefore the
probability distribution of the velocities is given by
\begin{align}
p(V_x,v_{x1},\ldots,v_{xN})dv_{x1}\cdots dv_{xN}dV_x \propto &\exp\left(-\frac{mv_{x1}^2+\cdots +
    mv_{xN}^2 +MV_x^2}{2k_BT}\right)\nonumber \\
&\qquad\times \delta(mv_{x1}+\cdots+mv_{xN}+MV_x) dv_{x1}\cdots dv_{xN}dV_x
\end{align}
where $\delta(\cdot)$ is the Dirac delta function. The probability
distribution of $V_x$ is then obtained by integrating over the $N$
particle velocities, 
\begin{equation}
p(V_x) \propto \int_{-\infty}^\infty dv_{x1}
\cdots\int_{-\infty}^\infty dv_{xN}\,\exp\left(-\frac{mv_{x1}^2+\cdots +
       mv_{xN}^2 +MV_x^2}{2k_BT}\right)\delta(mv_{x1}+\cdots+mv_{xN}+MV_x).
\end{equation}

To evaluate this integral we use the relation
\begin{equation}
  \delta(x)=\frac{1}{2\pi}\int_{-\infty}^\infty d\kappa \,\exp(i\kappa x).
  \label{eq:delta}
\end{equation}
Then
\begin{equation}
p(V_x) \propto \exp\left(-\frac{MV_x^2}{2k_BT}\right)\int_{-\infty}^\infty d\kappa
\exp(i\kappa MV_x)\left[\int_{-\infty}^\infty dv \,\exp\left(-\frac{mv^2}{2k_BT}+i\kappa mv\right)\right]^N.
\end{equation}
We use the result 
\begin{equation}
\int_{-\infty}^\infty du \exp(i\alpha u-\beta u^2) =2\int_0^\infty \cos(\alpha u)\exp(-\beta u^2)=
\left(\frac{\pi}{\beta}\right)^{1/2}\exp\left(-\frac{\alpha^2}{4\beta}\right),
\quad \beta>0.
\label{eq:expint}
\end{equation}
The integral inside the square brackets is then
\begin{equation}
\int_{-\infty}^\infty dv \,\exp\left(-\frac{mv^2}{2k_BT}+i\kappa
  mv\right)=\left(\frac{2\pi k_BT}{m}\right)^{1/2}\exp\left(-\half
  mk_BT\kappa^2\right).
\end{equation}
Thus
\begin{equation}
p(V_x) \propto \exp\left(-\frac{MV_x^2}{2k_BT}\right)\int_{-\infty}^\infty
d\kappa\,\exp\left(i\kappa MV_x-\half
  Nmk_BT\kappa^2\right).
\end{equation}
The integral can be evaluated using equation (\ref{eq:expint}), and
yields 
\begin{equation}
p(V_x) \propto \exp\left[-\frac{MV_x^2}{2k_BT}\left(1+\frac{M}{M_g}\right)\right].
\end{equation}
The mean-square velocity of the box along the $x$-axis is thus
\begin{equation}
  \langle V_x^2\rangle =\frac{k_BT}{M}\frac{M_g}{M+M_g}.
  \label{eq:main}
\end{equation}
This result agrees with equation (\ref{eq:en}) when the gas mass is
much larger than the box mass ($M_g\gg M$) and agrees with
(\ref{eq:mom}) when the gas mass is much less than the box mass. When
$M_g\ll M$, the argument leading to equation (\ref{eq:en}) is wrong
because it neglects the constraint imposed by momentum
conservation. When $M_g\gg M$ the argument leading to equation
(\ref{eq:mom}) is wrong because it neglects energy equipartition
between the gas and the box.

Although equation (\ref{eq:main}) was derived for a cubical box it
should hold for a box of arbitrary shape.

There is a simple heuristic derivation of this
result \citep{lt80}. Approximate the system as consisting of two bodies,
the box with mass $M$ and the gas with mass $M_g$. The reduced mass of
the two-body system is $\mu=MM_g/(M+M_g)$. In thermal equilibrium the
kinetic energy per degree of freedom is $\half k_BT$ so the relative
velocity $V_{x,\mathrm{rel}}$ between the two bodies satisfies
\begin{equation}
\langle V_{x,\mathrm{rel}}^2\rangle =\frac{k_BT}{\mu}.
\end{equation}
Since the center of mass is fixed, the one-dimensional velocity of the box is 
$V_x=M_g V_{x,\mathrm{rel}}/(M+M_g)$ so
\begin{equation}
\langle V_x^2\rangle =\frac{M_g^2}{(M+M_g)^2}\langle V_{x,\mathrm{rel}}^2\rangle=\frac{k_BT}{M}\frac{M_g}{M+M_g},
\end{equation}
the same as equation (\ref{eq:main}).

\section{The mean-square displacement of the free-floating box}

We can use similar methods to calculate the mean-square displacement
of the box. Assume that the center of mass of the system is at the
origin so $MX=-m\sum_{i=1}^N x_i$ where $x_i$ is the position of
particle $i$ and $X$ is the position of the center of mass of the
box. We assume that the box is cubical, with thin walls of length $L$
in each dimension.

Since we are only concerned with the position of the box, we can
integrate the probability distribution in phase space over the
velocities, so
\begin{equation}
  p(X,x_1,\ldots,x_N) \propto \delta(mx_1+\cdots+ x_N+
  MX)\prod_{i=1}^N\Theta(\half L+X-x_i)\Theta(\half L-X+x_i),
\end{equation}
where $\Theta(x)$ is the step function, equal to 1 if $x>0$ and 0 if
$x<0$. The delta function constrains the center of mass to remain at
the origin and the step functions constrain the particles to remain
inside the box. Using the identity (\ref{eq:delta}),
\begin{align}
   p(X,x_1,\ldots,x_N) &\propto \int_{-\infty}^\infty d\kappa\,
   \exp(i\kappa MX)\prod_{i=1}^N
   \int_{-L/2+X}^{L/2+X}dx_i\,\exp(im\kappa x_i) \nonumber \\
&\propto \int_{-\infty}^\infty d\kappa\,
   \exp[i\kappa X(M+M_g)]\left(\frac{\sin\half m L\kappa}{\kappa}\right)^N.
   \label{eq:intx}
 \end{align}
To evaluate the integral we assume that the gas mass $M_g=Nm$ is fixed
and let $N\to\infty$ or $m\to 0$. We replace the dummy variable $\kappa$ by $u\equiv\kappa/\surd N$. Then 
\begin{equation}
   p(X,x_1,\ldots,x_N) \propto \int_{-\infty}^\infty du\,
   \exp[iuN^{1/2}(M+M_g)] \left(\frac{\sin\half M_gLN^{-1/2}u}{u}\right)^N.
\label{eq:zzz}
\end{equation}
Since $|\sin x/x|$ is less than unity except near $x=0$,
$(\sin x/x)^N$ is negligible as $N\to\infty$ except near $x=0$. Here
it can be approximated by the first two terms in its Taylor series,
$\sin x/x=1-\frac{1}{6}x^2$, and we have
\begin{equation}
 \left(\frac{\sin\half M_g LN^{-1/2}u}{u}\right)^N\propto
 \left(1-\frac{M_g^2L^2u^2}{24N}\right)^N\to
 \exp\left(-\frac{M_g^2L^2u^2}{24}\right), 
\end{equation}
in which we have used the identity $(1-z/N)^N\to \exp(-z)$ as
$N\to\infty$. The integral in equation (\ref{eq:zzz}) can now be
evaluated using equation (\ref{eq:expint}), and we find
\begin{equation}
p(X) \propto \exp\left[-\frac{6N X^2(M+M_g)^2}{L^2M_g^2}\right].
\end{equation}
Note that as $N\to\infty$ this derivation converges to the correct
answer at a fixed value of $XN^{1/2}$ but not necessarily at a fixed
value of $X$ (obviously, $|X|$ cannot exceed $\half L M_g/(M+M_g)$,
the value it would have if all of the gas were on one wall of the
box). The mean-square displacement of the box from the original center
of mass is
\begin{equation}
\langle X^2\rangle =\frac{L^2}{12N}\left(\frac{M_g}{M+M_g}\right)^2.
\label{eq:disp}
\end{equation}
The factor $L^2/12$ is simply the variance of a uniform distribution
between $-\half L$ and $\half L$, given by
$\int_{-L/2}^{L/2} x^2dx/L$. For a box of general shape the factor
$L^2/12$ should be replaced by $\frac{1}{3}$ of the mean-square
distance of the box wall from its center of mass.

\section{Nuclear star clusters}

``Supermassive'' black holes with masses between $M_\bullet=10^6$ and
$M_\bullet=10^{10}$ solar masses are found in the centers of most
galaxies \citep{kh13}. The black holes are typically surrounded by
nuclear star clusters \citep{neu20}, systems of stars orbiting the
black hole. These are the densest stellar systems in the universe,
exceeding $10^7$ times the stellar number density in the solar
neighborhood. The mass of stars in the nuclear cluster can be
comparable to or even exceed the mass of the black hole. Because the
star cluster is usually much denser than the surrounding galaxy, it is
reasonable to approximate the cluster plus black hole as an isolated
system.

The dynamics and statistical mechanics of stars in a nuclear star
cluster differ from those of a gas in a box in several important
respects: (i) the stars are bound to the black hole because of their
own gravity and the gravity from the black hole, rather than being
confined in a box; (ii) relaxation occurs through gravitational
encounters, rather than collisions, similar to relaxation through
Coulomb encounters in a plasma; (iii) the mean free path is much
longer than the size of the system; (iv) even the oldest, densest
clusters have only lived for a few tens of relaxation times; (v) the
cluster cannot approach thermodynamic equilibrium, since it has
negative heat capacity \citep{bt08,bw76}.

The velocity and displacement of the black hole from the center of the
nuclear star cluster are accessible to observations, at least in
the nearest galaxies. The supermassive black hole at the center of the
Milky Way is nearly stationary: its velocity components
perpendicular to the line of sight are only $0.5\pm2.2\kms$ in the
Galactic plane and $0.9\pm0.8\kms$ normal to the plane \citep{reid20},
and the component along the line of sight \citep{genzel19} is
$3\pm2\kms$. The supermassive black hole in the nearest giant galaxy,
M31, is displaced from the center of the large-scale galaxy \citep{kb99}
by $0.07\pm0.01$ arcseconds or $0.23\pm0.03$ parsecs, although this
offset is probably due to the gravitational influence of the eccentric
stellar disk that orbits the black hole. The black hole in the giant
elliptical galaxy M87, which has been imaged by the Event Horizon
Telescope, appears to be offset from the photo-center of the galaxy by
$7\pm1$ pc \citep{batch10,blr11}.

The free-floating box provides a toy model that captures some but not
all of the dynamics that determines the Brownian motion of the black
hole due to its interactions with the surrounding stars, both within
the cluster and outside it. The nuclear star cluster is not in thermal
equilibrium, so the ``temperature'' or mean-square velocity of the
stars varies with radius, given roughly by the Kepler relation
$\langle v^2\rangle \simeq GM_\bullet/r$. If the number of stars in a small
radius range $dr$ is $dN$, and the mass of the star cluster is small
compared to the mass of the black hole, then equation (\ref{eq:main})
suggests that these contribute
$\langle V^2\rangle =\langle v^2\rangle m^2 dN/M_\bullet^2$ where $m$
is now the typical stellar mass. The mean-square velocity of the black
hole due to interactions with the cluster stars is then
\begin{equation}
\langle V^2\rangle \simeq \frac{Gm^2}{M_\bullet}\int \frac{dN}{r}.
\end{equation}
Similarly the mean-square displacement should be roughly
\begin{equation}
    \langle R^2\rangle\sim \frac{m^2}{M_\bullet^2}\int r^2\,dN, \quad Nm \lesssim M_\bullet.
\end{equation}
For the usual Bahcall--Wolf model of an equilibrium
cluster \citep{bw76}, $dN\sim r^{1/4}dr$, both the velocity and the
displacement are dominated by the effects of stars in the outer parts
of the cluster, and the mean-square velocity and radius of the black
hole are smaller than those of the stars in this region by a factor of
order $m^2N/M_\bullet^2$. A more quantitatively reliable approach, of
course, is to use N-body simulations \citep{lt80,chl02,reid20}.

\section{Summary}

We have described aspects of the Brownian motion of a free-floating
box containing an isothermal gas. The mean-square velocity and
displacement of the box are given by equations (\ref{eq:main}) and
(\ref{eq:disp}); the mean-square velocity is independent of the size
and shape of the box, and the mean-square displacement is independent
of the temperature. This system provides a toy model that describes
some aspects of the Brownian motion of a supermassive black hole in a
nuclear star cluster.  

We thank Walter Dehnen and Jean-Baptiste Fouvry for the discussions
that stimulated us to write this paper.

This work was supported in part by the Natural Sciences and
Engineering Research Council of Canada (NSERC), funding reference
number RGPIN-2020-03885; by the Black Hole Initiative at Harvard
University, which is funded by grants from JTF and GBMF; and by the
European Research Council (ERC) under the European Union's Horizon
2020 Programme for Research and Innovation ERC-2014-STG under grant
agreement No. 638435 (GalNUC).

\end{document}